\documentclass[lettersize,journal]{IEEEtran}
\usepackage{amsmath,amsfonts}
\usepackage{algorithmic}
\usepackage{algorithm}
\usepackage{array}
\usepackage[caption=false,font=normalsize,labelfont=sf,textfont=sf]{subfig}
\usepackage{textcomp}
\usepackage{stfloats}
\usepackage{url}
\usepackage{verbatim}
\usepackage{graphicx}
\usepackage{cite}
\usepackage{booktabs}
\usepackage{scalerel,amssymb}
\usepackage{xcolor}
\usepackage{bm}

\begin{document}

\title{Vaccinating Federated Learning for Robust Modulation Classification in Distributed Wireless Networks}


\author{Hunmin Lee$^{1}$, Hongju Seong$^{2}$, Wonbin Kim$^{3}$, Hyeokchan Kwon$^{4}$, and Daehee Seo$^{3}$%
\thanks{$^{1}$Hunmin Lee is with the Department of Computer Science and Engineering, University of Minnesota, Minneapolis, 55455, USA
{\tt\small lee03915@umn.edu}}
\thanks{$^{2}$Hongju Seong is with the Department of Computer Education, Sunchon National University, Suncheon, 57922, Republic of Korea 
{\tt\small labmen42@gmail.com}}
\thanks{$^{3}$Wonbin Kim and Daehee Seo are with the Department of Artificial Intelligence and Data Engineering, Sangmyung University, Seoul, 03016, Republic of Korea
{\tt\small wbkim29@smu.ac.kr, daehseo@smu.ac.kr}}
\thanks{$^{4}$Hyeokchan Kwon is with the Cyber Security Research Division of the Electronics and Telecommunications Research Institute (ETRI), Daejeon, 34129, Republic of Korea
{\tt\small hckwon@etri.re.kr}}
}

\markboth{IEEE Transactions on Cognitive Communications and Networking}%
{Lee \MakeLowercase{\textit{et al.}}: Vaccinating Federated Learning for Robust Modulation Classification in Distributed Wireless Networks}


\maketitle

\begin{abstract}

Automatic modulation classification (AMC) serves a vital role in ensuring efficient and reliable communication services within distributed wireless networks. Recent developments have seen a surge in interest in deep neural network (DNN)-based AMC models, with Federated Learning (FL) emerging as a promising framework. 
Despite these advancements, the presence of various noises within the signal exerts significant challenges while optimizing models to capture salient features. Furthermore, existing FL-based AMC models commonly rely on linear aggregation strategies, which face notable difficulties in integrating locally fine-tuned parameters within practical non-IID (Independent and Identically Distributed) environments, thereby hindering optimal learning convergence.
To address these challenges, we propose \textit{FedVaccine}, a novel FL model aimed at improving generalizability across signals with varying noise levels by deliberately introducing a balanced level of noise. This is accomplished through our proposed \textit{harmonic noise resilience} approach, which identifies an optimal noise tolerance for DNN models, thereby regulating the training process and mitigating overfitting. Additionally, FedVaccine overcomes the limitations of existing FL-based AMC models' linear aggregation by employing a split-learning strategy using structural clustering topology and local queue data structure, enabling adaptive and cumulative updates to local models. 
Our experimental results, including IID and non-IID datasets as well as ablation studies, confirm FedVaccine's robust performance and superiority over existing FL-based AMC approaches across different noise levels. These findings highlight FedVaccine's potential to enhance the reliability and performance of AMC systems in practical wireless network environments.

\end{abstract}


\section{Introduction} \label{sec1:intro}


\IEEEPARstart{O}{ver} the course of time, there has been a rapid evolution in wireless communication technologies, particularly in their applications integrated with the Internet of Things (IoT), providing substantial benefits to global-wide users~\cite{chettri2019comprehensive, sikimic2020overview}. Notably, the infusion of Artificial Intelligence (AI) technology into wireless communication has significantly contributed to enhancing the efficiency of various communication systems, encompassing network optimization~\cite{chang2021survey}, resource management~\cite{baccour2022pervasive}, Multiple-Input and Multiple-Output (MIMO) system operation~\cite{9896861}, enhancing network security~\cite{zaman2021security}, and optimizing Quality of Service (QoS)~\cite{xu2021artificial}. 

The incorporation of AI in wireless networks, particularly in the domain of Automatic Modulation Classification (AMC) tasks, has led to significant performance improvement in the modulation recognition systems~\cite{huynh2021automatic}.
Given the widespread utilization of AMC techniques in practical scenarios, such as cellular networks, Wi-Fi systems, satellite communication, radar systems, and other wireless technologies, the integration of AI technology in AMC has brought high-performance and effective AMC schemes across diverse conditions in the wireless IoT network~\cite{jdid2021machine}.
AMC technology contends with a multitude of signals emanating from diverse user devices dispersed across varied environments. Within this distributed framework, the conventional paradigm of centralized learning presents notable drawbacks in terms of privacy concerns and resource constraints, including large communication bandwidth costs and storage expenses associated with transmitting and storing locally curated datasets to a central server.

Federated learning (FL) emerges as a suitable paradigm for addressing those constraints, primarily due to its intrinsic characteristics that preserve privacy, alleviate communication overhead, and substantially reduce storage utilization~\cite{FL}. The decentralized nature of FL enables local model training on edge devices, eliminating the necessity to transmit raw data to a central server. 
The adaptability of the FL framework within the heterogeneous nature of user data enhances operational effectiveness across distributed IoT systems, concurrently promoting cost efficiency and fortifying the system against faults. Furthermore, the continuous learning capability after deployment inherent in FL proves pivotal for time-sensitive applications, as evidenced by its application in modulation classification within dynamic communication landscapes~\cite{abdel2021survey, peng2017modulation, jdid2021machine}. Therefore, the manifold advantages of FL illustrate a necessary framework for addressing AMC challenges in wireless networks.


However, in the context of a distributed wireless system, where data is collected from diverse devices under certain user conditions, the impact of noise becomes particularly pronounced. 
The performance of AMC models is heavily dependent on the quality of the input datasets, making the inherent noise in wireless signals a critical challenge to their resilience. Existing research in AMC~\cite{mod_fed0, mod_fed1, mod_fed2_fedbkd, mod_fed3_automatic, abdel2021survey, huynh2021automatic, jdid2021machine, peng2021survey} has predominantly evaluated model effectiveness under specific noise conditions, typically quantified by Signal-to-Noise Ratio (SNR). These studies consistently demonstrate that models trained on high SNR data perform well, while those exposed to low SNR data struggle.

This emphasis on high SNR data aligns with the conventional wisdom that low-noise signals simplify the training of deep neural network (DNN)-based modulation decoders by enabling the extraction of clear, distinguishable features. However, this focus inadvertently fosters a bias, suggesting that high SNR conditions are universally optimal for training DNN models. This perspective risks promoting overfitting, as models trained exclusively on high SNR data may fail to generalize across diverse noise environments.
In real-world applications, wireless communication systems often encounter a broad range of noise levels, resulting in significant noise variance that AMC models must contend with. The prevailing focus on high SNR conditions in training does not adequately address this variability, thereby undermining the generalizability and robustness of AMC models in practical, noise-prone environments. Addressing this gap is crucial for developing more resilient and adaptable AMC systems capable of maintaining performance across varying and unpredictable noise conditions.

Furthermore, recent research has proposed the utilization of FL in AMC models, aiming to harness the benefits of FL methodologies within distributed environments~\cite{mod_fed0, mod_fed1, mod_fed2_fedbkd, mod_fed3_automatic, siriwardana2023federated, majeed2020blockchain, wang2020distributed, wei2021architecture}. Prior studies on FL-based AMC models have predominantly revolved around addressing the challenges posed by non-IID (Independent and Identically Distributed) environments within distributed settings. However, existing works often narrowly target singular non-IID issues, especially a class imbalance problem~\cite{mod_fed0, siriwardana2023federated, qi2022collaborative}, overlooking the myriad of other non-IID complexities inherent in distributed datasets. These complexities encompass variations in dataset volume, statistical distributions across distributed clients, incongruent features, and SNR discrepancies.

Moreover, the current FL-based AMC models predominantly rely on a linear aggregation approach, which exhibits notable limitations in seamlessly integrating locally optimized parameters. This process often leads to information loss during aggregation, thereby compromising the efficacy of collaborative learning, particularly within non-IID environments. Importantly, this challenge is not unique to FL-based AMC models but is also pervasive in conventional Federated Averaging (FedAvg)-based methodologies~\cite{zhang2021survey, wen2023survey}. Addressing these limitations is paramount for advancing the effectiveness and scalability of FL-based AMC models in real-world distributed settings.
To summarize, the existing constraints in FL-based AMC classification can be delineated as follows:

\begin{itemize}
    \item The enduring challenge posed by diverse noise sources in modulation signals highlights the critical importance of implementing effective noise management strategies in distributed wireless networks.
    
    \item The common practice of exclusively evaluating models based on specific SNR values may foster a bias towards the belief that consistently high SNR levels are necessary for AMC model training, potentially resulting in overfitting issues in real-world scenarios characterized by diverse SNR ranges.
    
    \item The existing aggregation process in FL-based AMC models, relying on linear-based parameter aggregation, faces challenges in effectively integrating models trained under non-IID conditions, thereby constraining its performance in heterogeneous environments.
    
\end{itemize}

To address these challenges, we introduce a novel FL framework \textit{FedVaccine}. This framework is grounded on two fundamental principles. Firstly, inspired by the concept of vaccination in the medical domain, FedVaccine incorporates a controlled noise exposure strategy during DNN model training to foster resilient modulation classification performance across diverse noise levels. Leveraging our harmonic noise resilience methodology, we systematically explore the optimal noise tolerance within signals, thereby achieving a delicate balance between dataset robustness and model regularization to mitigate overfitting issues. We comprehensively investigate the impact of noise tolerance of the DNN-based AMC model, revealing that models trained with balanced levels of noise exhibit superior performance over those trained solely with high SNR signals, thus enhancing the model's resilience and generalizability.

Secondly, diverging from conventional linear aggregation methods employed in FL, FedVaccine adopts a split learning approach. This strategy entails partitioning multiple participant local parameter sets into distinct clusters and subsequently integrating intra-cluster models while cumulatively updating across inter-cluster iterations. Moreover, we incorporate an adaptive queue data structure to mitigate bias within non-IID settings, thereby addressing practical memory constraints within local devices.
This nuanced approach fine-tunes the global model to preserve pre-trained parameter attributes, thereby minimizing information loss during the integration process. 

Our extensive experimentation, spanning a wide range of noise levels and three prevalent non-IID scenarios, demonstrates the remarkable performance enhancement of FedVaccine compared to existing FL-based AMC models. These results underscore FedVaccine's efficacy in achieving accurate and resilient modulation classification, thereby making significant contributions to the advancement of wireless communication technology.
In summary, the contributions of our work are summarized as follows:

\begin{itemize}
    \item \textit{\textbf{Noise-Resilient Training Strategy}}: We introduce a harmonic noise resilience approach that achieves balanced noise tolerance, regularized model training, and enhances the generalizability of modulation classification across diverse noise levels. 
    
    \item \textit{\textbf{Introduction of FedVaccine}}: We propose a novel Federated Learning framework named FedVaccine, designed to address practical non-IID issues and enhance optimization for robust modulation classification performance.  

    \item \textit{\textbf{Comprehensive Experimental Validation}}: We conduct extensive experiments and ablation studies to evaluate FedVaccine's performance and demonstrate its efficiency and practical applicability.
 
\end{itemize}



This manuscript is structured as follows. Section~\ref{sec2:related_works} provides an exploration of the footprints of AMC studies in distributed IoT networks, denoising schemes in AMC, and prior FL approaches. In Section~\ref{section3}, the preliminary concepts directly relevant to our work are outlined. In Section~\ref{section4}, we present our methodologies of harmonic noise resilience and FedVaccine.
In Sections \ref{section 5} and \ref{section 6}, thorough experiments are undertaken to validate the efficacy of FedVaccine across multiple datasets and scenarios. 
Section~\ref{sec7} explains the real-world significance and novelties of our study, as well as the limitations and future works. Finally, we conclude our study in Section~\ref{sec8}, summarizing our works.

\begin{figure*}
    \centering
    \includegraphics[width=\linewidth]{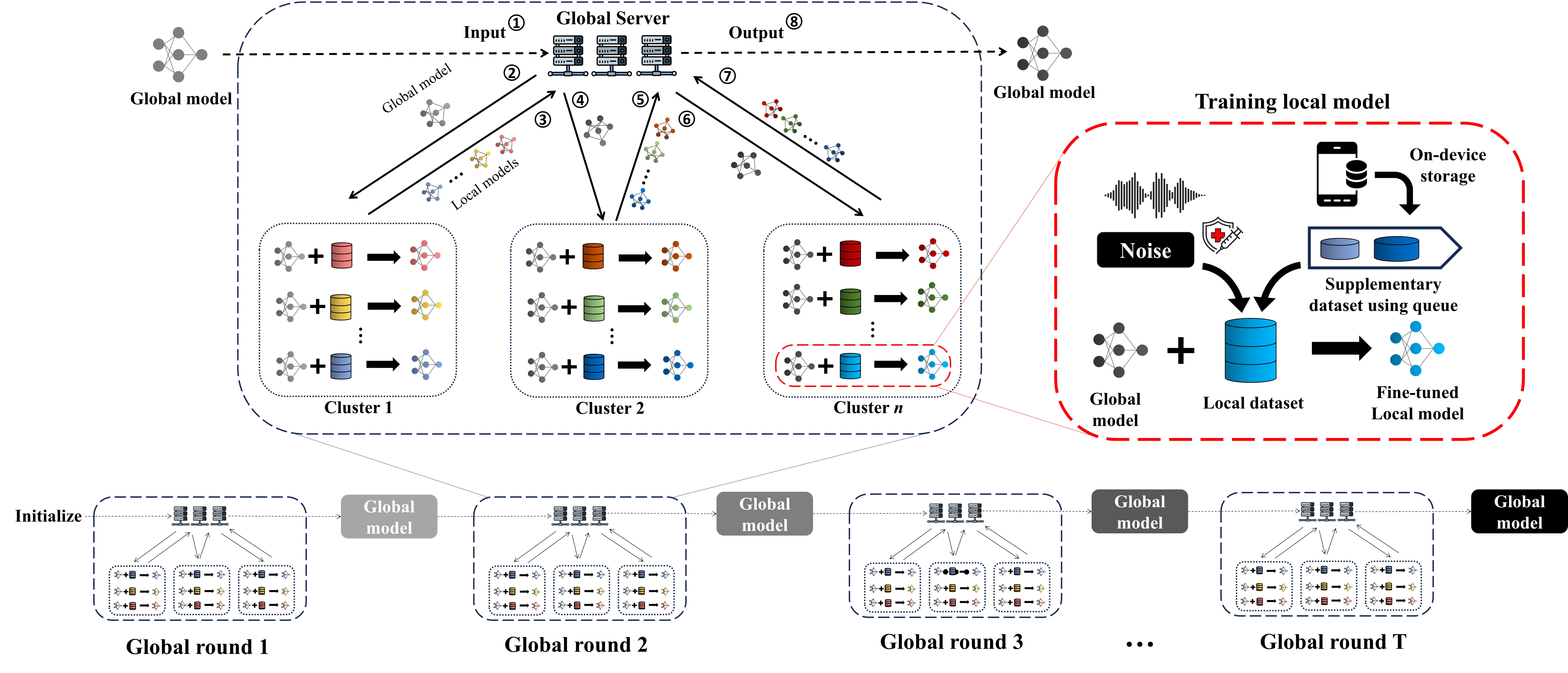}
    \caption{Architectural overview of FedVaccine. FedVaccine involves deliberate accommodation of a controlled level of noise and the incorporation of supplementary datasets through a queue structure aimed at mitigating bias in non-IID scenarios. Utilizing these locally fine-tuned models, the global model undergoes iterative updates through cluster-wise units, minimizing the information loss during the aggregation stage.}
    \label{fig:teaser}
\end{figure*}

\section{Related Works} \label{sec2:related_works}

\subsection{Modulation Classification in Distributed IoT Networks}

AMC technology is a critical component of modern wireless systems, providing a range of advantages that include improved adaptability, optimized spectral efficiency, guaranteed QoS, and support for cognitive radio functionalities~\cite{abdel2021survey}. By allowing wireless IoT networks to function efficiently in dynamic and challenging environments, AMC technology enhances the ability of communication systems to coordinate, optimize, and maintain consistent, reliable performance in the face of fluctuating conditions inherent in contemporary wireless frameworks.

Advancements in machine learning have significantly impacted the field of AMC, leading to the widespread adoption of machine learning frameworks within this domain, as highlighted in previous studies~\cite{hazza2013overview, jdid2021machine}. Various machine learning methodologies have been applied, including Support Vector Machines (SVM)~\cite{park2008automatic, sengur2009multiclass}, Bayesian networks\cite{liu2016modulation, krayani2021automatic}, random forests~\cite{zhang2017method, zhao2019low}, and ensemble learning approaches~\cite{liu2017research}. These methods have proven effective in accurately identifying modulation types by leveraging features intrinsic to the models. Building on the success of these traditional machine learning techniques, DNN architectures have gained prominence in AMC tasks~\cite{peng2021survey, huynh2021automatic}. In particular, Convolutional Neural Networks (CNNs) have become a favored choice due to their ability to efficiently extract both local and global features within the spatial domain, resulting in superior performance in AMC applications~\cite{o2016convolutional, peng2017modulation, zhou2019robust, zhang2020automatic}.

Moreover, recognizing the significance of temporal attributes inherent in modulation signals, there has been a concerted exploration into extracting temporal dynamics for effective classification. Recurrent Neural Network (RNN)-based models have thus been applied and developed to capture temporal dependencies, employing architectures such as Gated Recurrent Unit (GRU)~\cite{utrilla2020gated}, Long Short-Term Memory (LSTM) network~\cite{zhang2020automatic}, and transformer model~\cite{cai2022signal}.
Furthermore, diverse deep learning paradigms have been harnessed to enhance performance and construct scalable, efficient architectures within the AMC domain. These include methodologies such as transfer learning~\cite{wang2020transfer}, reinforcement learning~\cite{zhou2022electromagnetic}, adversarial learning~\cite{bu2020adversarial}, and meta learning strategies~\cite{vuorio2019multimodal}, all contributing to augmenting the capabilities of AMC systems.

\subsection{Denoising in Modulation Classification}  \label{sec2-2}

As advanced machine learning paradigms were applied in AMC tasks, the quality of the signal datasets holds paramount importance during model training. As the widely known expression \textit{Garbage-in, Garbage-out} represents, it is widely recognized that the persistence of unexpected noise within signals has presented a longstanding challenge throughout the history of wireless signal processing. 
Notably, the term `noise' encompasses a spectrum of definitions across various domains. In the context of this study, noise refers to an unforeseen disturbance detected at the receiver, originating from either internal or external sources. 

Within the AMC domain, numerous studies have been dedicated to addressing the noise inherent within signals. Bagga \textit{et al.,}~\cite{bagga2012study} was one of the pioneering AMC studies considering SNR conditions, introducing a model utilizing wavelet transform and a statistical parametric-based method to build an AMC model.
Moreover, as the usage of machine learning models evolves, subsequent studies have predominantly focused on developing robust models based on machine learning approaches~\cite{wu2017robust, abdel2021survey, jdid2021machine}. 
More recently, there has been a surge in leveraging DNN models for AMC across varying SNR conditions~\cite{hu2019deep, han2021automatic, khan20223d, abdel2021survey, huynh2021automatic}. Hu \textit{et al.,}~\cite{hu2019deep} proposed a modulation classifier utilizing LSTM, demonstrating superior performance when SNR exceeds 10dB and outperforming Expectation Maximization-based algorithms across diverse SNR ranges.
Han \textit{et al.,}~\cite{han2021automatic} transformed time-domain signals into frequency-based features through a combination of CNN and stacked autoencoder, employing the Probabilistic Neural Network (PNN) model for AMC across multiple SNR ranges.
Furthermore, Khan \textit{et al.,} \cite{khan20223d} designed an AMC model based on a 3D CNN architecture under various noise environments, including additive white Gaussian noise and Rayleigh/Rician channel, leveraging spatiotemporal information for robust model training.
Collectively, diverse model architectures have been proposed to mitigate noise under varying conditions, aiming to construct a resilient classifier capable of handling noisy signal environments effectively.



\subsection{Federated Learning for Modulation Classification} \label{sec2-3:related_works}

Federated Learning (FL)~\cite{FL} has gained widespread recognition as an apt framework for distributed environments, harnessing collective knowledge from participating local devices to facilitate collaborative learning. Likewise, FL has garnered considerable interest within the domain of modulation classification technology, seeking to establish an adaptive framework tailored to the distributed IoT environment~\cite{mod_fed1, majeed2020blockchain, wei2021architecture, shi2021signal, zhang2022deep, mod_fed0, siriwardana2023federated, qi2022collaborative, mod_fed2_fedbkd}. 
Shi \textit{et al.},~\cite{mod_fed1} leveraged FL in the AMC field, which observed the impact of training the DNN model over different scenarios across edge models, including various training dataset volumes, different SNR, varying numbers of edge clients within the distributed environment. 
Inspired by this, diverse studies were proposed that applied FL in AMC task, which can be narrowed down to two large categories: \textit{enhancing privacy}~\cite{majeed2020blockchain, wei2021architecture, shi2021signal, zhang2023attacking}, and \textit{achieving optimization under non-IID conditions}~\cite{mod_fed0, siriwardana2023federated, qi2022collaborative, mod_fed2_fedbkd}.


\subsubsection{Security in AMC Federated Learning}
To ensure privacy, Majeed \textit{et al.},~\cite{majeed2020blockchain} leveraged the blockchain framework in FL-based AMC to enhance security levels across participants in wireless IoT-edge systems. Wei \textit{et al.,}~\cite{wei2021architecture} experimentally explored diverse attack scenarios in FL in the AMC setting, comparing the performance variance using multiple deep learning models using a public dataset. Additionally, Shi \textit{et al.,}~\cite{shi2021signal} employed a differential privacy scheme, preserving performance and enhancing the privacy level during FL operation. Apart from the countermeasures for adversarial attacks, Zhang \textit{et al.,}~\cite{zhang2023attacking} proposed a new poisoning attack method for modulation recognition FL framework in an IoT environment. Although FL has significantly increased the privacy level compared to centralized learning, this study implies that it still involves vulnerability to adversarial attacks and malicious activities.

\subsubsection{Non-IID Optimization in AMC Federated Learning}
In the domain of FL, it is well recognized that non-IID datasets present significant challenges to achieving optimal convergence. Recent research efforts have increasingly focused on addressing the non-IID characteristics commonly encountered in distributed environments, particularly within the AMC domain. A prominent challenge in this context is the class imbalance problem, which often arises in non-IID classification tasks in distributed learning settings. To mitigate this issue, Wang \textit{et al.}~\cite{mod_fed0} proposed FedeAMC, a method that addresses class imbalances by utilizing a balanced cross-entropy function to effectively distribute class type weights. Similarly, Siriwardana \textit{et al.}~\cite{siriwardana2023federated} employed data augmentation techniques to enhance the performance of FL-based AMC, particularly in low Signal-to-Noise Ratio (SNR) and non-IID scenarios, effectively addressing class imbalance concerns. Additionally, the Federated Imbalanced Learning (FIL) approach~\cite{qi2022collaborative} was introduced to tackle class imbalance, demonstrating superior performance compared to traditional FedAvg methods in such environments. Furthermore, FedBKD~\cite{mod_fed2_fedbkd} proposed a model that creates synthetic datasets using variational autoencoders on the server side, combined with bidirectional knowledge distillation techniques to train local models. This approach effectively mitigates heterogeneity from both data and model perspectives within the distributed learning framework.


\section{Preliminaries} \label{section3}

\subsection{Modulation Classification and Noise}

\subsubsection{Automatic Modulation Classification}

Modulation classification is a fundamental component in modern wireless communication systems, enabling the identification and categorization of modulation schemes within received signals. Its primary goal is automatic and accurate recognition of modulation types without human intervention, ensuring reliable communication across diverse user environments~\cite{abdel2021survey}. Through analysis of signal features like constellation, spectral characteristics, and temporal properties, AMC algorithms classify signals into predefined types such as amplitude modulation (AM), frequency modulation (FM), phase shift keying (PSK), and quadrature amplitude modulation (QAM). Previous studies~\cite{jdid2021machine, park2008automatic, sengur2009multiclass} have successfully extracted relevant features from signals and mapped them to modulation classes using various machine learning techniques. These technologies are crucial for adaptive radio applications, promoting efficient spectrum utilization and robust communication in dynamic environments.

\subsubsection{Preliminaries of Noise}

In the wireless communication domain, noise has been a long-lasting challenge stemming from internal and external factors. Internally generated noise, including Gaussian noise~\cite{rappaport2024wireless}, equipment noise~\cite{rappaport2024wireless}, impulse noise~\cite{selim2020noma}, and synchronization noise~\cite{mani2018architecture}, originates within communication systems, partly within the control of station operators. Mitigating internal noise involves strategies~\cite{rappaport2024wireless, nguyen2004cmos, zhang2021learning} such as low-noise amplifiers, filtering methodologies, error correction models, shielding techniques, and design optimizations aimed at enhancing the system's resilience against noise interference. Conversely, external noise presents a more daunting challenge as its origins lie beyond station operators' influence, characterized by its unpredictable nature and persistence as a perturbation regardless of station condition. 
Common sources of external noise include frequency interference~\cite{njoku2005global}, multipath fading~\cite{vaigandla2021study}, and shadowing~\cite{rappaport2024wireless}. 

These sources, contributing to diminishing signal characteristics, are significant concerns for optimizing distributed system operation. The unpredictable nature of noise components, coupled with SNR variations, presents challenges in training accurate modulation classification models. 
Previous studies on AMC using DNN approaches have emphasized the importance of clean, high SNR signals while discerning modulation types~\cite{hazza2013overview, jdid2021machine, peng2021survey, huynh2021automatic, o2016convolutional, peng2017modulation, zhou2019robust, zhang2020automatic, utrilla2020gated, zhang2020automatic, cai2022signal, mod_fed0, mod_fed1, mod_fed2_fedbkd, siriwardana2023federated, qi2022collaborative}, highlighting the necessity of denoising datasets. As wireless system deployments continue to expand, understanding and mitigating noise's impact on AMC becomes integral to advancing reliable and adaptive modulation classification techniques for evolving wireless communication systems.

\subsection{Federated Learning-based AMC in IoT Network}

\subsubsection{Federated Learning}   FL is a decentralized machine learning approach where model training is conducted collaboratively across multiple participant devices without centralizing raw data~\cite{FL}. 
Let $\mathbf{w}_{i}$ denote the local model with index $i$, trained locally using resources and datasets $\mathbb{D}$ specific to each local node, with a task-based loss function $L(\cdot)$ as detailed in equation~\eqref{eq:local_train}. The objective function in equation~\eqref{eq:local_train_obj} guides the iterative update process with time $t$ for optimizing $\mathbf{w}$ towards minimizing the local loss function.

\begin{equation}
    \mathbf{w}^{(t+1)} = \mathbf{w}^{(t)} - \eta \nabla L(\mathbf{w}^{(t)}_{i}, \mathbb{D}_{i}) 
    \label{eq:local_train}
\end{equation}

\begin{equation}
    \lim_{t \rightarrow T} \mathbf{w}^{(t)}: \rightarrow \min_{\mathbf{w}} L(\mathbf{w}, \mathbb{D}) 
    \label{eq:local_train_obj}
\end{equation}

Subsequently, the fine-tuned local models $\mathbf{w}_{\forall i}$ from all local nodes undergo aggregation at the global server, synchronized in a timely manner. The aggregation process, depicted in equation~\eqref{eq:FL_aggregation}, constructs a global model $\mathbf{W}$ through element-wise matrix aggregation across each layer, where $q$ signifies the weights assigned to each model based on the dataset volume.

\begin{equation}
    \mathbf{W} = \frac{1}{|\forall i|} \sum_{\forall i} q_{i}\mathbf{w}_{i}
    \label{eq:FL_aggregation}
\end{equation}

This global model is then redistributed back to the participating devices, facilitating the update of their local models. This cycle initiates successive rounds of equations from \eqref{eq:local_train} to \eqref{eq:FL_aggregation}, iteratively refining the global model. 

\subsubsection{Modulation Classification in FL Environment}

In distributed wireless environments, the application of FL to modulation classification emerges as a noteworthy approach. This involves the collaborative participation of numerous IoT devices, each equipped with signal communication functions. 
The deployment of the FL framework in modulation classification within distributed wireless environments affords several merits. Foremost is the commitment to data privacy, as sensitive signal information remains decentralized on individual devices, mitigating concerns related to data security and regulatory compliance. Furthermore, the collaborative nature of FL leverages the collective knowledge of diverse participant user devices, thereby enhancing the accuracy of modulation classification. This decentralized approach proves particularly advantageous in scenarios where centralized methods face impracticalities, either due to the scale of IoT devices or concerns pertaining to communication latency. Thus, FL-based modulation classification stands out as a promising paradigm for optimizing the efficiency, accuracy, and privacy aspects of wireless communication systems within distributed environments.

\subsection{Non-IID Problem in Distributed Environment} \label{sec3-3}

In a distributed environment within wireless communication systems, signals traverse diverse regions or channels and are subject to disparate environmental conditions, interference, and noise levels. The characteristics of regional noise profiles have significant variability, inducing dissimilarities in received signals.
In modulation classification, wherein the objective is to discern the modulation type of a received signal, these fluctuations in regional noise and other conditions present challenges, giving rise to non-IID problems.
This section defines and categorizes prevalent non-IID scenarios encountered in FL-based modulation classification tasks within a distributed wireless environment, where we will implement these scenarios in the experiment section (Sections ~\ref{section 5} and \ref{section 6}).

\subsubsection{Case 1. Class Imbalance} \label{sec:non-iid-1}

The issue of class imbalance is a prevalent challenge under non-IID conditions~\cite{mod_fed0, siriwardana2023federated, qi2022collaborative}. This problem occurs when class instances are unevenly distributed, as illustrated in equation~\eqref{eq:non-iid case1} where $D(\cdot)$ represents distribution, $c$ is a class, $U(\cdot)$ indicates uniform distribution, and $\mathbf{y}$ represents ground-truth class.
Such imbalance reduces the model's sensitivity to minority classes and introduces significant bias during the fine-tuning process, thereby impeding the development of a generalizable performance in FL environments.

\begin{equation}
    D(\mathbf{y}) \nsim U(0, |\forall c|-1) \ s.t. \ \mathbf{y} \in \mathbb{D}
    \label{eq:non-iid case1}
\end{equation}

\subsubsection{Case 2. Dataset Volume Imbalance} \label{sec:non-iid-2}

The dataset volume imbalance arises from an uneven distribution of data samples among local users.
Devices or sensors responsible for data collection may contribute varying volumes of local datasets, resulting in some users generating significantly more samples than others. This imbalance poses challenges for machine learning models trained on such datasets, as it can lead to biases favoring specific users with larger datasets, potentially skewing global model performance in FL. This imbalance is quantified in equation~\eqref{eq:non-iid case2}, where $n(\cdot)$ represents a number of samples, $i$ and $j$ denote arbitrary local users.

\begin{equation}
    n(\mathbb{D}_{i}) \not\approx n(\mathbb{D}_{j}) \ where \ i \ne j
    \label{eq:non-iid case2}
\end{equation}

\subsubsection{Case 3. Feature Variance} \label{sec:non-iid-3}

The feature variance issue reflects the variations from the inherent features across the unique local datasets. Let $\mathbf{X}$ be the input signals with classes $\mathbf{y}$, where $\mathbb{D} \ni (\mathbf{X}, \mathbf{y})$, and $f(\mathbf{X})$ indicates the feature extractor using input $\mathbf{X}$. We define the non-IID case 3 as follows:

\begin{equation}
    f(\mathbf{X}_{i}) \not\approx f(\mathbf{X}_{j}) \ where \ i \ne j  
    \label{eq:non-iid case3}
\end{equation}

In the context of modulation signals, the modulation scheme itself remains consistent across different devices; however, variability is introduced by external factors, such as noise affecting the original modulated signal. In this study, we applied a specific SNR range that varies across different regions, with the implementation details provided in Section~\ref{sec6-1}.

\section{Methodology}  \label{section4}

\subsection{Problem Definition} 

Prior to presenting our methodology, we define the prevailing problems in the modulation classification domain within distributed user environments.
Our investigation is centered on two key challenges. Firstly, we explore the inherent noise complexities in practical modulation signals and highlight the gap between conventional AMC studies and real-world settings.
Secondly, we explain the persistent issue of non-IID data distribution and inefficiencies in conventional FL models during optimization in parameter aggregation.




\subsubsection{Balancing Noise and Signal in Modulation Classification}



In spite of the well-established notion that modulated signals characterized by low noise facilitate the effective extraction of discernible features by DNN models for modulation classification~\cite{zhang2022deep}, real-world signals frequently exhibit noise stemming from various sources of interference. This phenomenon invariably leads to a noticeable deterioration in model performance during the practical inference phase, necessitating the formulation of effective strategies to reduce the disparity between real-world test inference and the preparatory phase of model training. Traditional methodologies for noise reduction, as discussed in Section~\ref{sec2-2}, typically entail key challenges. It encompasses the risk of information loss during denoising procedures and imposing substantial computational overhead on lightweight user devices during real-time operations. Recent endeavors have geared towards the adoption of AI-based techniques, encompassing the extraction of salient feature representations, the harnessing of advanced machine learning models for efficacious feature learning, or the assumption of constrained environmental conditions, such as specific SNRs. Despite their commendable contributions towards enhancing classification accuracy, prior AMC schemes remain susceptible to the intrinsic noise prevalent in signals, constituting a foundational impediment necessitating redress. 

Our investigation takes a new approach by prioritizing the equilibrium between authentic signal components and noise within modulation signals. Diverging from conventional methodologies that train DNN classifiers using modulated signals with an arbitrary range of SNR, our approach endeavors to pinpoint an optimal noise bandwidth intrinsic to the signal spectrum, thereby enabling the DNN classifier to achieve generalizable performance across a diverse array of incoming signals characterized by varying SNRs. 

Notably, our proposed methodology, \textit{harmonic noise resilience} approach, orchestrates the equilibrium of extracted features between noise and genuine signal components, while concurrently regulating the training process to delineate a robust decision boundary. By identifying a balanced noise level that maximizes model performance, our approach aims to facilitate harmonious interaction between noise and signal to enhance the generalizability of handling signals with diverse SNRs. We introduce our harmonic noise resilience methodology in Section~\ref{sec4-2}.

\subsubsection{Federated Learning Design for AMC}


In distributed computing environments, FL presents a notable advantage by enabling the collaborative aggregation of knowledge dispersed among locally trained DNN models, all converging towards a common task objective.
Recent investigations~\cite{mod_fed0, mod_fed1, mod_fed2_fedbkd, mod_fed3_automatic, siriwardana2023federated} underscore the efficacy of FL models in AMC, thereby enhancing practicality through distributed modeling.
Despite the advancements, prior studies have predominantly focused on an isolated and singular non-IID issue, particularly class imbalance, whereas the challenges inherent in a distributed modulation classification environment are manifold, as elucidated in the preceding Section~\ref{sec3-3}. To achieve real-world deployment readiness, it is imperative to delve further into and address additional challenges that align with practical scenarios.

Beyond the limited exploration of non-IID problems, conventional FedAvg-based AMC methodologies encounter significant hurdles during the aggregation phase of locally trained parameters. Specifically, the rudimentary linear aggregation of parameter collections fails to facilitate optimal integration across heterogeneously fine-tuned parameter sets tailored to their respective datasets.
In fact, this challenge extends beyond the AMC domain, encompassing various domains leveraging FL models.

To address these challenges, we propose a new FL model, \textit{FedVaccine}, designed to iteratively refine the global model. Our approach aims to alleviate the influence of non-IID distributions while rectifying the shortcomings associated with linear aggregation, achieved through the iterative re-training of the global model using cluster configuration. Moreover, we merge the harmonic noise resilience method into FedVaccine, enhancing the generalizability. FedVaccine design is delineated in Section~\ref{sec4-3}.



\subsubsection{Notation}

Before introducing our methodology, a compilation of frequently utilized notations is presented in Table~\ref{tab:notation}.

\begin{table}[t]
\centering
\caption{Notation Table}
\begin{tabular}{cc|cc}
    \toprule
    \textbf{Notation} & \textbf{Description} & \textbf{Notation} & \textbf{Description}\\
    \midrule
    $\mathbf{x}$  & input sample  & $\mathbf{y}$  & ground truth label \\ 
    $T$           & global epoch  & $t$           & local epoch \\  
    $\mathbf{t}$  & time          & $i,j$         & index  \\
    $\mathbf{W}$  & global model  & $\mathbf{w}$  & local model \\
    $\theta$      & SNR threshold & $L(\cdot)$    & Loss function \\
    $\mathbb{D}$  & dataset       & $\kappa$      & error \\
    $\theta(\cdot)$ & noise signal & $s(\cdot)$   & modulation signal \\
    $\mathbf{Q}$  & queue         & $\vartheta$   & memory capacity \\
    $\ell$  & layer         & $\delta$      & number of trained dataset \\
    $D(\cdot)$    & distribution  & $R(\cdot)$    & noise source \\
    a $\oplus$ b  & append b to a & a $\ominus$ b & remove element b from a \\
    \bottomrule
\end{tabular}
\label{tab:notation}
\end{table}

\subsection{Harmonic Noise Resilience} \label{sec4-2}

In this section, we introduce a methodology for determining the optimized noise level within the training dataset for modulation classification, namely the `Harmonic noise resilience' approach. 
Let $\mathbf{W}$ be an initialized parameter for the DNN-based modulation classification model, yielding a modulation prediction $\hat{\mathbf{y}}$ through the function $f(\mathbf{W}, \mathbf{X})$. The prediction is evaluated with ground truth $y$ using a cross-entropy function in equation~\eqref{eq:loss_crossentropy}, where $i$ signifies the sample index and $j$ represents the class index, respectively.
Using a predefined function in equation~\eqref{eq:local_train_obj}, it iteratively updates the $\mathbf{W}$ by leveraging equation~\eqref{eq:loss_crossentropy}.

\begin{equation}
    L(y_{ij}, \hat{y}_{ij}) = -\frac{1}{|\mathbb{D}|}\sum_{\forall i}\sum_{\forall j}y_{ij}\log(\hat{y}_{ij})
    \label{eq:loss_crossentropy}
\end{equation}

Here, the training dataset $\mathbf{X} \ni \mathbf{x} (i.e., \mathbb{D} \ni (\mathbf{X}, \mathbf{y}))$, can be factorized into original signal $s(\mathbf{t})$ and noise signal $\epsilon(\mathbf{t})$ over time $\mathbf{t}$ using equation~\eqref{eq:signal_noise_range}.

\begin{equation}
    \mathbf{x} = s(\mathbf{t}) + \epsilon(\mathbf{t}) 
    \label{eq:signal_noise_range}
\end{equation}

In equation~\eqref{eq:signal_noise_range}, $\epsilon(t)$ consists of an arbitrary range of noise levels, with $R_{k}(\mathbf{t})$ denoting an arbitrary noise composed of trigonometric function signal from source $k$ and $\sum_{k} R_{k}(\mathbf{t})$ indicating the combined noise forming $\epsilon(\mathbf{t})$, as described as follows:

\begin{equation}
    \epsilon\mathbf{t} = \sum_{k} R_{k}(\mathbf{t}), \ e.g. \ R(\mathbf{t}) = A\sin_{k}(\omega \mathbf{t} + \phi)
    \label{eq:noise}
\end{equation}

where $A$ represents amplitude, $\omega$ is the angular frequency, and $\phi$ is the phase angle.
Next, the SNR of $\mathbf{x}$ is defined by equations \eqref{eq:SNR1} and \eqref{eq:SNR2}, with the time interval $[0,\varsigma]$. Using these two equations, we measure the quality of signal $\mathbf{x}$ with respect to noise. 

\begin{equation}
    SNR(\mathbf{x}) = \frac{P_{signal}}{P_{noise}} \\ 
    \label{eq:SNR1}
\end{equation}

\begin{equation}
    P_{signal} = \frac{1}{\varsigma}\int_{0}^{\varsigma}|s(\mathbf{t})|^{2} d\mathbf{t}, \ \ P_{noise} = \frac{1}{\varsigma}\int_{0}^{\varsigma}|\epsilon(\mathbf{t})|^{2} d\mathbf{t}
    \label{eq:SNR2}
\end{equation}

The following equation~\eqref{eq:training_data_define} demonstrates the selection of the training dataset using a threshold $\theta$, which filters noise ranges based on SNR values, specifically retaining those higher than $\theta$. By default, this includes the highest SNR range.

\begin{equation}
    \mathbf{X} \ni 
    \begin{cases}
     \mathbf{x} \qquad \text{if} \ SNR(\mathbf{x}) > \theta \\
     \emptyset \qquad \text{otherwise} \\
    \end{cases}     
    \label{eq:training_data_define}
\end{equation}

With the prepared dataset, parameter $\mathbf{W}$ is fine-tuned using $\mathbf{X}$ filtered with $\theta$, aiming to minimize loss using equation~\eqref{eq:min_threshold}.

\begin{equation}
    \arg\min_{\mathbf{t}} [\bigcup_{\forall \mathbf{t}} L(\mathbf{y}^{(\mathbf{t})}_{\theta}, (\mathbf{W}^{(\mathbf{t})}_{\theta}, \mathbf{X}^{(\mathbf{t})}_{\theta}))]:\rightarrow \mathbf{W}^{(\mathbf{t})}_{\theta}    \label{eq:min_threshold}
\end{equation}

Subsequently, an evaluation function $E(\cdot)$ is defined to compute the ratio of correctly classified elements using the test dataset $\hat{\mathbb{D}}_{test}$, as depicted in equation~\eqref{eq:evaluation}, where $N$ represents the total number of samples in $\hat{\mathbb{D}}_{test}$, and $I(\cdot)$ denotes an indicator function that returns 1 if the condition inside the parentheses is true, otherwise 0.

\begin{equation}
    E(\mathbf{W}^{(\mathbf{t})}_{\theta}, \hat{\mathbb{D}}_{test}) = \frac{\sum^{N}_{i=1}I(\hat{y}_{i} = y_{i})}{N}
    \label{eq:evaluation}
\end{equation}

Finally, our objective function is defined in equation~\eqref{eq:obj_harmonic_noise}, finding $\theta$ that returns the highest performance across various SNR values.

\begin{equation}
    \arg \max_{\theta} [\bigcup_{\forall \theta} E(\mathbf{W}^{(\mathbf{t})}_{\theta}, \hat{\mathbb{D}}_{test})]
    \label{eq:obj_harmonic_noise}
\end{equation}


\subsection{FedVaccine Model} \label{sec4-3}



In this section, we introduce a new FL framework \textit{FedVaccine}.
The foundational architecture of FedVaccine is in the iterative update progression by clusters, facilitating the transfer of acquired knowledge from a clustered set of models to the subsequent cluster. In contrast to conventional FL models'~\cite{zhu2021federated, zhang2021survey} linear aggregation approach, where the parameters of all participants jointly merge and generate a representative model, our approach of sequential cluster-wise integration aims to mitigate information loss during the aggregation of knowledge. 
Specifically, the local models were fundamentally fine-tuned with local datasets, with personalized adaptation within the unique local environment. However, during integration, simply merging models in a linear fashion dilutes the inherent capability across heterogeneous parameters. This information loss becomes much more pronounced in non-IID scenarios, where local attributes are highly distinguishable and explicit. 
Therefore, our sequential update approach strategy is simple yet offers significant advantages, particularly in non-IID scenarios, where it effectively addresses challenges arising from parameter heterogeneity and subsequent discordance during the aggregation process.
Additionally, the weighted aggregation method allows for normalizing and balancing the contributions of models based on their significance. This equilibrium is particularly crucial in scenarios where certain local models possess more pertinent or accurate information for specific tasks, serving as an effective strategy in practical non-IID scenarios. 

Moreover, by employing a threshold parameter $\theta$ during the dataset preprocessing stage, we optimize the classification performance by selecting an appropriate SNR range to curate the most effective training dataset to impart resilience to adverse noises. The selection of a minimum threshold range aims to balance a reasonable variance of SNR to adaptively train models, serving as a regularization strategy that enhances the generalizability of models within diverse noise levels.

Finally, our proposed framework incorporates a queue data structure $\mathbf{Q}$ for individual local devices, allowing each device to manage a designated memory capacity resource for the storage of supplementary data. To ensure the model maintains its currency and adapts to evolving performance requirements, the First-In-First-Out (FIFO) method is implemented within the queue. This involves the storage of newly acquired datasets while systematically removing outdated ones.
During the dataset storage process, a condition is enforced to approximate the class label distribution of the stored dataset to a ground truth uniform distribution \(\mathbf{D}\) with an error term \(\kappa\), as denoted in equation~\eqref{eq:A}, where \(D(\mathbf{y})\) signifies the distribution of the label vector \(\mathbf{y}\). 

\begin{equation}
    \centering
    \begin{aligned}
        D(\mathbf{y}) \approx \mathbf{D} + \kappa, \ s.t. \ \mathbf{D} \approx \mathbf{y} \sim U(a,b), \\ 
        where \ a \leq y \leq b \ and \ y \in \mathbf{y} \ \ \ \ \ \ \
    \end{aligned}
    \label{eq:A}
\end{equation}

Additionally, upon the acquisition of fresh datasets, the Jensen-Shannon (JS) Divergence \(D(P||Q)\) between the label distribution of the acquired dataset and our ground truth \(\mathbf{D}\) is computed in equations~\eqref{eq:B} and~\eqref{eq:C}. 

\begin{equation}
    D_{KL}(D(\mathbf{y})||\mathbf{D}) = \sum_{\forall i}(D(\mathbf{y}_{i}) \times log(\frac{D(\mathbf{y}_{i})}{\mathbf{D}_{i}}))
    \label{eq:B}
\end{equation}

\begin{equation}
    \begin{aligned}
        D(P||Q) = \frac{1}{2}D_{KL}(P||\frac{(P+Q)}{2})+ \frac{1}{2}D_{KL}(Q||
\frac{(P+Q)}{2}) \\
where \ P \leftrightarrow D(\mathbf{y}), \mathbf{D} \leftrightarrow Q \ \ \ \ \ \ \ \ \ \ \ \ \ \ \ \
    \end{aligned}
\label{eq:C}
\end{equation}

The resulting disparity \(\hat{D}\) informs the identification of specific elements \(\mathbf{q}\) to be removed, as illustrated in equation~\eqref{eq:D}, where \(\mathbf{Q}.\text{pop}(n)\) represents the indicator function that pops the element $n$ from $\mathbf{Q}$. 

\begin{equation}
    \begin{aligned}
        \hat{D} = \mathbf{D} - D(P||Q), \ s.t. \ \hat{D} \leftrightarrow \mathbf{q} \subset \mathbf{Q} \\ 
        \mathbf{Q}.\text{pop}(n): pop \ n \ from \ \mathbf{Q}, \ where \ n \in \mathbf{q} 
    \end{aligned}
    \label{eq:D}
\end{equation}

This mechanism allows focused preservation of new input data, significantly mitigating non-IID attributes and effectively reducing the non-IID effect while training for modulation classification tasks.
The proposed FedVaccine is elucidated in detail in the algorithm~\ref{alg1}.


\begin{algorithm}
\caption{: FedVaccine Algorithm}
\textbf{Input}: Local datasets $\mathbb{D}_{i}^{\mathbf{t}} \ni \mathbf{x}_{i}, \mathbf{y}_{i}$, Local Queue $\mathbf{Q}_{i}$ \\
\textbf{Output}: Global model $\mathbf{W}^{(T)}$
\begin{algorithmic}[1]
    \STATE \vspace{0.1cm} Initialize all participant client $i$'s model $\textbf{w}^{(T=0)}_{i}$
    \STATE \vspace{0.1cm} Initialize global model $\mathbf{W}^{(T=0)}$ in central server
    \STATE \textbf{for} \textit{global epoch} $T$ = 1,2, ..., $\mathbf{T}$ \textbf{do}
    \STATE \hspace{0.5cm} Run the following for all clients in parallel
    \STATE \hspace{1cm} Curate new local dataset $\mathbb{D}_{i}^{(T)}$
    \STATE \hspace{1cm} $\mathbf{Q}_{i}$.insert($\mathbb{D}_{i}^{(T)}$)
    \STATE \hspace{1cm} $\mathbf{z}_{i}$ = SNR($\mathbf{x}_{i}$)
    \STATE \hspace{1cm} \textbf{for} j = 1,2, ..., $n(\mathbf{z}_{i})$ \textbf{do}
    \STATE \hspace{1.5cm} \textbf{if} $z_{(i,j)} < \theta$ \textbf{then}
    \STATE \hspace{2cm} $\mathbb{D}^{(T)}_{i} \ominus z_{(i,j)} $  
    \STATE \hspace{1cm} \textbf{if} $T > \vartheta $ \textbf{then}
    \STATE \hspace{1.5cm} $\mathbf{Q}_{i}$.pop($\mathbb{D}_{i} - D(D(\textbf{y}_{i}^{(T-\vartheta)}) || \mathbf{D})$)
    \STATE \hspace{1cm} \textbf{if} $T > 1 $ \textbf{then}
    \STATE \hspace{1.5cm} $\mathbb{D}^{(T)}_{i} \oplus \mathbf{Q}_{i}$

    \STATE \hspace{0.5cm} \textbf{for} cluster $c$ = 1,2, ...,C \textbf{do}
    
    \STATE \hspace{1cm} Train $\mathbf{w}_{i}^{(T)}$ using $\mathbb{D}^{(T)}_{i}$ with \textit{b} mini-batches
    
    \STATE \hspace{1cm} Collect ($\mathbf{w}_{i}^{(T)}$, $\delta_{i}$) to central server
    \STATE \hspace{1cm} $\mathbf{W}_{\ell}^{(T)} = \frac{\sum_{i}}{n(\forall i)}(1 - \frac{\delta_{i}}{\sum_{\forall i} \delta_{i}}) \mathbf{W}_{\ell}^{(T-1)} + \frac{\delta_{i}}{\sum_{\forall i} \delta_{i}} \mathbf{w}_{(i,\ell)}^{(T)}$
        
    \STATE \hspace{1cm} Broadcast $\mathbf{W}^{(T)}$ to all clients in cluster $c$
    \STATE \textbf{return} $\mathbf{W}^{(T)}$
    \end{algorithmic}
    \label{alg1}
\end{algorithm}


\section{Experiment 1: Harmonic Noise Resilience} \label{section 5}



In this section, we implement the process defined in section~\ref{sec4-2} and report the corresponding results after conducting a thorough analysis to derive optimal $\theta$ for robust generalization within the AMC model.

\subsection{Setting} \label{sec5-1}

In the initial phases of our analysis, we evaluate the classification performance of two representative DNN models for processing spatial and temporal features: CNN and GRU. This assessment is concentrated on a model trained with an SNR reduction strategy, aiming to systematically assess the influence of both the degree and volume of noise within the training data.
For each model, we adopt the pre-designed architectures proposed by O’Shea \textit{et al.}~\cite{o2016convolutional} for CNN and Hong \textit{et al.}~\cite{hong2017automatic} for GRU, specifically tailored for the modulation classification task. 
In CNN, the padding scheme employed within the first Conv2D layer facilitated the preservation of the initial shape of the feature map, whereas the second Conv2D layer did not retain paddings.
These baseline models underwent a training process based on a grid search on $\theta$, incorporating signal data within a specified range of SNR. 

\subsubsection{Dataset Setting} \label{sec5-1-2}   In this experiment, the RML2016.10a dataset~\cite{o2016radio} curated by DeepSig was employed. 
RML2016.10a encompasses a modulation dataset generated using GNU Radio, featuring 11 modulation types (comprising 8 digital and 3 analog) across a range of SNR ratios from -20 to 18 dB, with increments of 2 dB. The dataset comprises a total of 220,000 samples, and their modulation classes present in RML2016.10a include 8PSK, AM-DSB, AM-SSB, BPSK, CPFSK, GFSK, PAM4, QAM16, QAM64, QPSK, and WBFM. 

To ensure a comprehensive evaluation, training and test datasets were randomly shuffled and divided in an 8:2 ratio, adhering to the specified search space range. Notably, the reduction of the search space of $\theta$ by a decrement of 2 was applied on the lower SNR side, aligning with the consensus that higher SNR values tend to yield more favorable learning outcomes (\textit{e.g.}, -20 $\sim$ 18, -18 $\sim$ 18, -16 $\sim$ 18, ..., 16 $\sim$ 18, 18). The list of $\theta$ is denoted as follows.

\begin{equation}
    \{-20, -18, ..., \theta, ..., 18|-20 \leq \theta \leq 18, \ where\ \theta \div 2 = 0\}
    \label{eq:threshold_candidate}
\end{equation}

\subsubsection{Hyperparameter Setting}   The hyperparameter configuration involved 500 epochs, an Adam optimizer, a batch size of 400, a learning rate set at 0.001, and a ReLu activation function before the decision-making layer. Importantly, the uniformity of training and test dataset volumes was maintained throughout the SNR reduction process. 
This uniformity was achieved by randomly selecting a quantity equivalent to the number of data instances with an SNR value of only 18, representing the minimum range set. To mitigate the impact of this randomness, the training process was repeated four times for each $\theta$.
All the experiments were conducted using CPU i9-12900KS, 32GB RAM, and GPU machines with Nvidia GeForce RTX 3070Ti and 3080Ti equipped with 8GB and 16GB VRAM, respectively.

\subsection{Results} \label{sec5-2}

\begin{figure*}
    \centering
    \includegraphics[width=\linewidth]{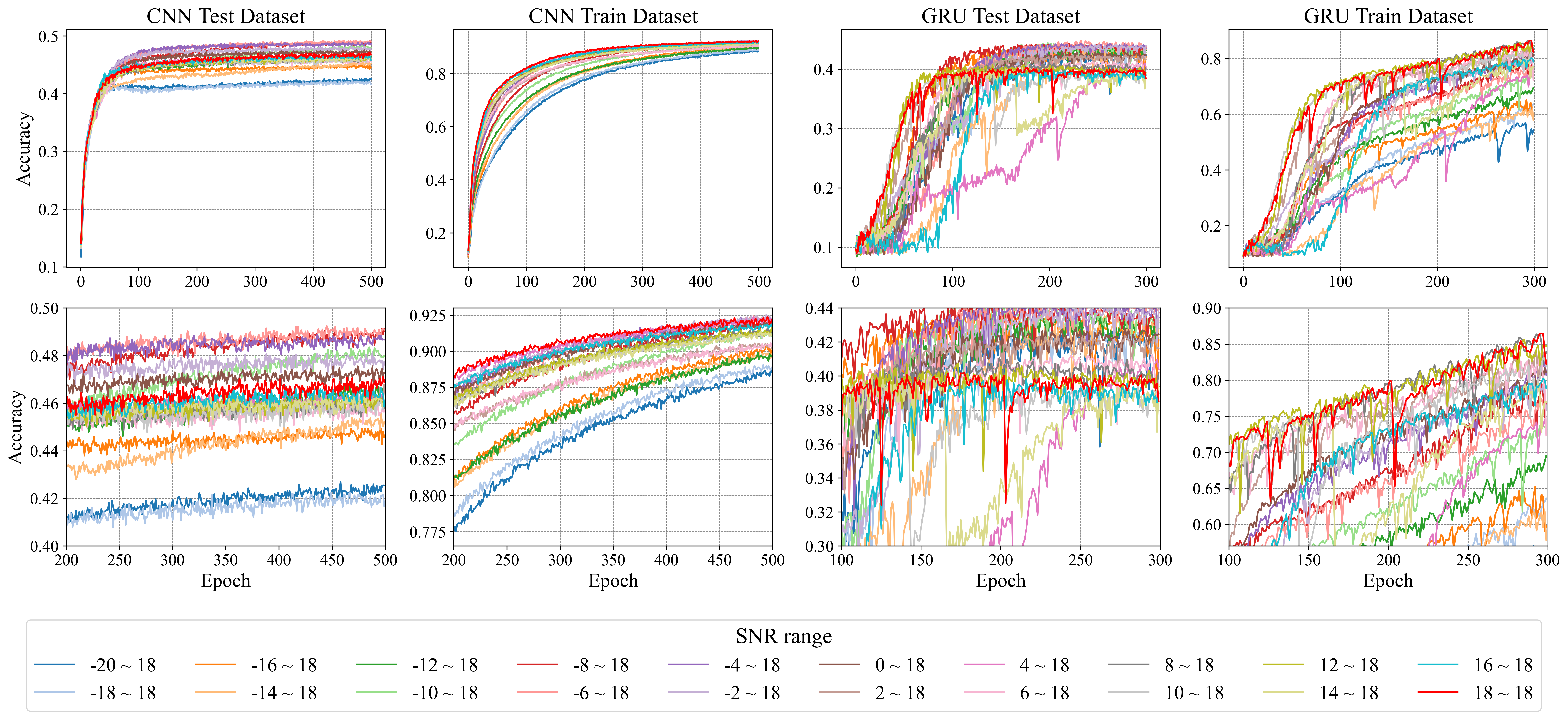}
    \caption{The training and test outcomes of CNN~\cite{o2016convolutional} and GRU~\cite{hong2017automatic} models are presented. The figures in the initial row depict the original results, while the corresponding enlarged versions of each column in the first row are displayed in the second row.}
    \label{fig:exp1}
\end{figure*}

\begin{table}[]
\centering
\caption{The test accuracy outcomes are presented across a varied range of SNR, with the value (x.x) denoting the standard deviation observed after conducting the training process four times in two different machines for each SNR range. Here, the SNR range is equivalent to $\theta \sim$ 18.}
\begin{tabular}{ccccc}
\hline
 & \multicolumn{2}{c}{CNN Accuracy (\%)}   & \multicolumn{2}{c}{GRU Accuracy (\%)}  \\ \hline
SNR (dB)      & Test data     & Train data  & Test data   & Train data  \\ \hline
-20 $\sim$ 18 &  40.98 (2.0)  & 74.73 (1.2) & 42.82 (0.3) & 57.03 (6.7) \\
-18 $\sim$ 18 &  40.54 (2.9)  & 75.52 (1.1) & 43.25 (0.4) & 63.62 (5.4) \\
-16 $\sim$ 18 &  43.42 (3.2)  & 77.68 (1.4) & 43.76 (0.4) & 65.21 (4.8) \\
-14 $\sim$ 18 &  42.79 (2.7)  & 77.42 (0.7) & 43.57 (0.8) & 61.85 (6.1) \\
-12 $\sim$ 18 &  44.32 (1.8)  & 77.99 (1.1) & 43.74 (0.3) & 69.56 (5.7) \\
-10 $\sim$ 18 &  45.28 (0.6)  & 80.49 (0.3) & 44.41 (0.1) & 73.70 (3.8) \\
-8 $\sim$ 18  &  \textbf{46.85 (0.5)}  & 81.93 (0.4) & \textbf{44.75 (0.1)} & 79.59 (3.9) \\
-6 $\sim$ 18  &  46.67 (0.3)  & 83.37 (0.1) & 44.65 (0.2) & 77.41 (2.4) \\
-4 $\sim$ 18  &  46.78 (0.2)  & 83.68 (0.3) & 44.35 (0.1) & 82.69 (1.7) \\
-2 $\sim$ 18  &  46.08 (0.7)  & 84.25 (0.4) & 44.15 (0.2) & 79.68 (1.6) \\ 
 0 $\sim$ 18  &  45.45 (0.4)  & 83.97 (0.7) & 42.86 (0.6) & 81.67 (2.1) \\ 
 2 $\sim$ 18  &  44.40 (1.4)  & 81.96 (1.6) & 42.39 (0.3) & 83.90 (3.4) \\ 
 4 $\sim$ 18  &  44.58 (0.5)  & 85.00 (0.1) & 40.99 (0.3) & 74.50 (3.5) \\ 
 6 $\sim$ 18  &  44.11 (0.2)  & 82.13 (1.0) & 41.26 (0.4) & 83.81 (1.0) \\  
 8 $\sim$ 18  &  44.21 (0.8)  & 83.50 (0.7) & 41.07 (0.2) & 86.42 (3.3) \\ 
 10 $\sim$ 18  & 44.35 (0.7)  & 83.55 (0.8) & 39.85 (0.6) & 83.06 (6.9) \\ 
 12 $\sim$ 18  & 44.40 (0.6)  & 83.93 (0.6) & 40.01 (0.4) & 85.67 (1.9) \\ 
 14 $\sim$ 18  & 44.34 (0.9)  & 83.39 (0.3) & 40.00 (0.1) & 79.54 (4.9) \\ 
 16 $\sim$ 18  & 44.75 (0.3)  & 84.66 (0.2) & 39.89 (0.2) & 80.26 (1.5) \\ 
 18            & 44.89 (0.6)  & \textbf{85.37 (0.4)} & 39.42 (0.4) & \textbf{86.50 (1.7)} \\ 
\hline
\end{tabular}
\label{Tab:one_model_snr_acc}
\end{table}

The maximum test and training results of CNN and GRU models are reported in Table~\ref{Tab:one_model_snr_acc}, accompanied by a visual representation of learning convergence in Fig.~\ref{fig:exp1}. In Fig.~\ref{fig:exp1}, the figures in the first row depict the overall convergence, while the second row exhibits an enlarged version of each corresponding figure above.
The performance comparison of CNN and GRU reveals similar outcomes, with CNN displaying a smoother convergence, while GRU exhibits a relatively unstable trajectory in its learning curve. 

\subsection{Discussion} \label{sec5-3}

Following the general consensus, the accuracy of training data peaks at the highest SNR (18 dB). However, it is notable the best test accuracy occurs when the signal is randomly mixed with noises within the SNR range of $-8$ to $18$ ($\theta = -8$), as highlighted in bold in Table~\ref{Tab:one_model_snr_acc}. This discrepancy suggests that models trained exclusively with high SNR may overfit compared to models trained with noise-embedded data, emphasizing that signals of high quality do not consistently yield an effective learning strategy.
Furthermore, the training accuracy in CNN exhibits a linearly proportional pattern to SNR, indicating that optimal training quality is achieved when learning representations from clean data with high signal quality. In terms of test accuracy, the threshold $\theta$ range of -8 $\sim$ -4 proves to be an effective range for training and classifying the given test dataset. Conversely, training and test dataset performance significantly declines when the signal involves an SNR range below -16.

This finding highlights that rather than exclusive reliance on a low-noise dataset, incorporating partially perturbed data with a specific noise level proves effective. The observed superior and robust performance patterns in both spatial-based CNN and temporal-based GRU models reveal a noteworthy phenomenon of \textit{harmonic noise resilience}, showcasing its generalizability across diverse DNN approaches. This suggests that a holistic approach, integrating noise-embedded data alongside instances with high SNR, may significantly enhance the robustness and versatility of DNN models tailored for AMC tasks.

\begin{figure*}
    \centering
    \includegraphics[width=\linewidth]{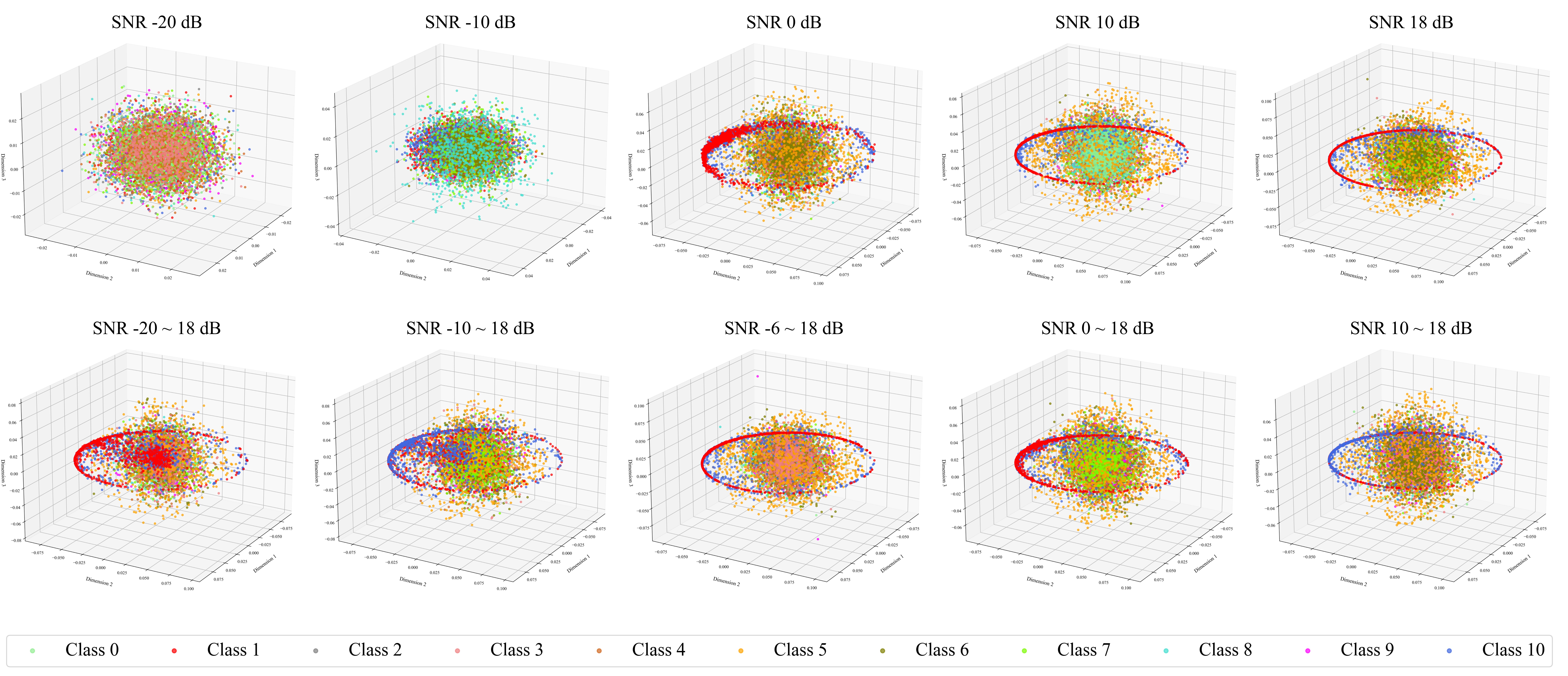}
    \caption{Visualization result after reducing feature dimensions through PCA. The figures arranged in the initial row depict the PCA outcomes corresponding to discrete SNR, while those in the second row illustrate the PCA results associated with the combined SNR range.}
    \label{fig:pca_result}
\end{figure*}

\subsection{Feature Analysis}

To further examine the influence exerted by noise on DNN models within the input signals, we translate the learned representations from each search space into a low-dimensional feature space. This enables visual exploration of the structural aspects of the data distribution, illustrating an interpretation of how the model captures the acquired representation.
Principle Component Analysis (PCA) is employed to reduce the dimensionality of each signal to three components. Fig.~\ref{fig:pca_result} depicts the results, with the figures in the first row sequentially representing signals with discrete SNR values of -20, -10, 0, 10, and 18. Simultaneously, the figures in the second row sequentially display the PCA results of signals with an SNR range of -20 to 18, -10 to 18, -6 to 18, 0 to 18, and 10 to 18.

Remarkably, the signals with discrete SNR values begin to distinctly reveal structural patterns across the 11 modulation classes, starting from 0 SNR value. 
The most distinguishing factor is that the components of classes 1 and 10 encircle the main cluster of points, with the ring-shaped configuration gradually becoming more vivid and aligning as the SNR increases, minimizing the variance. 
In the combined SNR, despite the involvement of a low degree of SNR, the results consistently display the ring, indicating the preservation of separable features along the dimensions. A comparison between PCA results of signals with SNR ranges -20 $\sim$ 18 and 10 $\sim$ 18 reveals a disparity in the ring, particularly the absence of class 10 in the first figure. This suggests that data points of class 10 (Wideband FM; WBFM) are substantially affected by low SNR, whereas class 1 (Amplitude Modulation with Double Sideband; AM-DSB) is comparatively less affected.
This observation aligns with the general knowledge of modulation, where WBFM may be more susceptible to noise due to its wider bandwidth and potential vulnerability to frequency deviations caused by noise. On the other hand, AM-DSB may exhibit a degree of resilience to low SNR, given its primary involvement with variations in amplitude rather than frequency.

This visualization analysis serves to underscore that even in the presence of perturbation among signals of high quality, discernible features are retained within principle components. As elucidated in the preceding sections \ref{sec5-2} and \ref{sec5-3}, this phenomenon of harmonic noise resilience imparts supplementary advantages, particularly in the selection of an optimal SNR range, enhancing the model's capability to capture meaningful representations within the dataset.

\section{Experiment 2: FedVaccine} \label{section 6}

\subsection{Setting} \label{sec6-1}

In the second experiment, we investigate the effectiveness of FedVaccine in comparison to existing FL models and alternative learning paradigms across both IID and various non-IID scenarios using two public datasets.

\subsubsection{Dataset}   During our experiments, we additionally employed RML2016.10b~\cite{o2016radio} dataset. 
RML2016.10b is also widely recognized as a standard benchmark for tasks involving modulation recognition through machine learning models, which is an extended version of RML2016.10a, encompassing a larger dataset comprising 1,200,000 modulation samples. Similar to RML2016.10a (see Section~\ref{sec5-1-2}), it spans the identical SNR ratio range while excluding a specific modulation class, AM-SSB, having 10 modulation class types.

\subsubsection{Model Setting}   Likewise to the previous experiments, the CNN model proposed by \cite{o2016convolutional} serves as our baseline local model architecture with identical hyperparameter protocols in \cite{o2016convolutional} except for training epochs. 
Here, the training comprised 10 local epochs, with a subsequent aggregation of over 100 global epochs. Within our distributed environment, we established the participation of 10 local clients. The training and test datasets were partitioned randomly with a 9:1 split ratio, with the local datasets iteratively sampled from the training data pool, each comprising 1000 samples. The queue size per local was set to 1500, allocating storage capacity to store 1500 samples, and the cluster size was set to 2, incorporating five models per cluster.



\subsubsection{Non-IID Scenario 1} 
In the initial non-IID scenario, we emulate the class imbalance issue across the distributed environment. The ratio of each class label in the local datasets is randomly selected, with the sampling process carried out independently for each local dataset and repeated in every global round. The random selection is done within the range of 0 to 100\%, where the number of samples is set to 1000.

\subsubsection{Non-IID Scenario 2} 
The second scenario introduces variability in the dataset volume across local devices. Similar to the first scenario, we assign random probabilities ranging from 0 to 100\% within the 1000 samples in each local and every global epoch, representing the ratio of preserving the original dataset. This probabilistic allocation is performed independently for each local dataset and is reiterated in every global round.

\subsubsection{Non-IID Scenario 3}
The final scenario is to allocate heterogeneous and random feature attributes across local datasets. In this setting, we randomly select a single SNR value and allocate the dataset within that SNR across local clients in each global epoch, where the number of samples is maintained between 400 and 600. 
Moreover, for all non-IID scenarios, the queue size per local was extended with 500 samples. 
This scenario aims to measure the performance of feature variance that may typically occur in the real world, where the SNR statistics may be biased and differ across the local environment.


\subsection{Comparison Models}

To validate the efficacy of our FedVaccine, we incorporate different learning paradigms and various FL models to comprehensively compare the performance across three aforementioned non-IID settings. 

\subsubsection{Global Learning}
Global Learning (GL) is a standard end-to-end learning process where we collect all the local datasets into the central server. In each local client, $N$ data samples were randomly collected in each non-IID case within the benchmark RML dataset and transmitted their datasets to the server for 100 global communication rounds, having $N$ (samples) $\times$ 10 (locals) $\times$ 100 (global rounds) = $N \times 1,000$ samples and training them in a global CNN model.

\subsubsection{Centralized Learning}
Centralized Learning (CL)~\cite{asad2021federated} is a framework that follows global learning in a distributed environment. The central server collects the local datasets in each global communication round, and the global model is trained in each communication round, constantly updating the model with new datasets.

\subsubsection{Distributed Learning}
The distributed learning (DistL) paradigm~\cite{asad2021federated} holds an environment similar to FL, whereas the DistL does not aggregate the local models but iteratively trains them with locally generated datasets across the global round without any communications across the distributed clients. 

\subsubsection{Federated Learning}
In our comparison of FL models, we assessed a total of nine models, including FedVaccine. Specifically, we focused on models with architectures that do not involve the sharing of information directly among the participating local clients. The selected models comprised FedAvg~\cite{FL}, FedSGD~\cite{FL}, FedProx~\cite{li2020federated}, FedBN~\cite{li2021fedbn}, FedMD~\cite{li2019fedmd}, FedPer~\cite{arivazhagan2019federated}, FedBKD\cite{mod_fed2_fedbkd}, FedDistill~\cite{jiang2020federated}, and FedSL~\cite{abedi2023fedsl}.


\subsection{Result in IID Environment} \label{sec:6-3}

\begin{figure*}
    \centering
    \includegraphics[width=\linewidth]{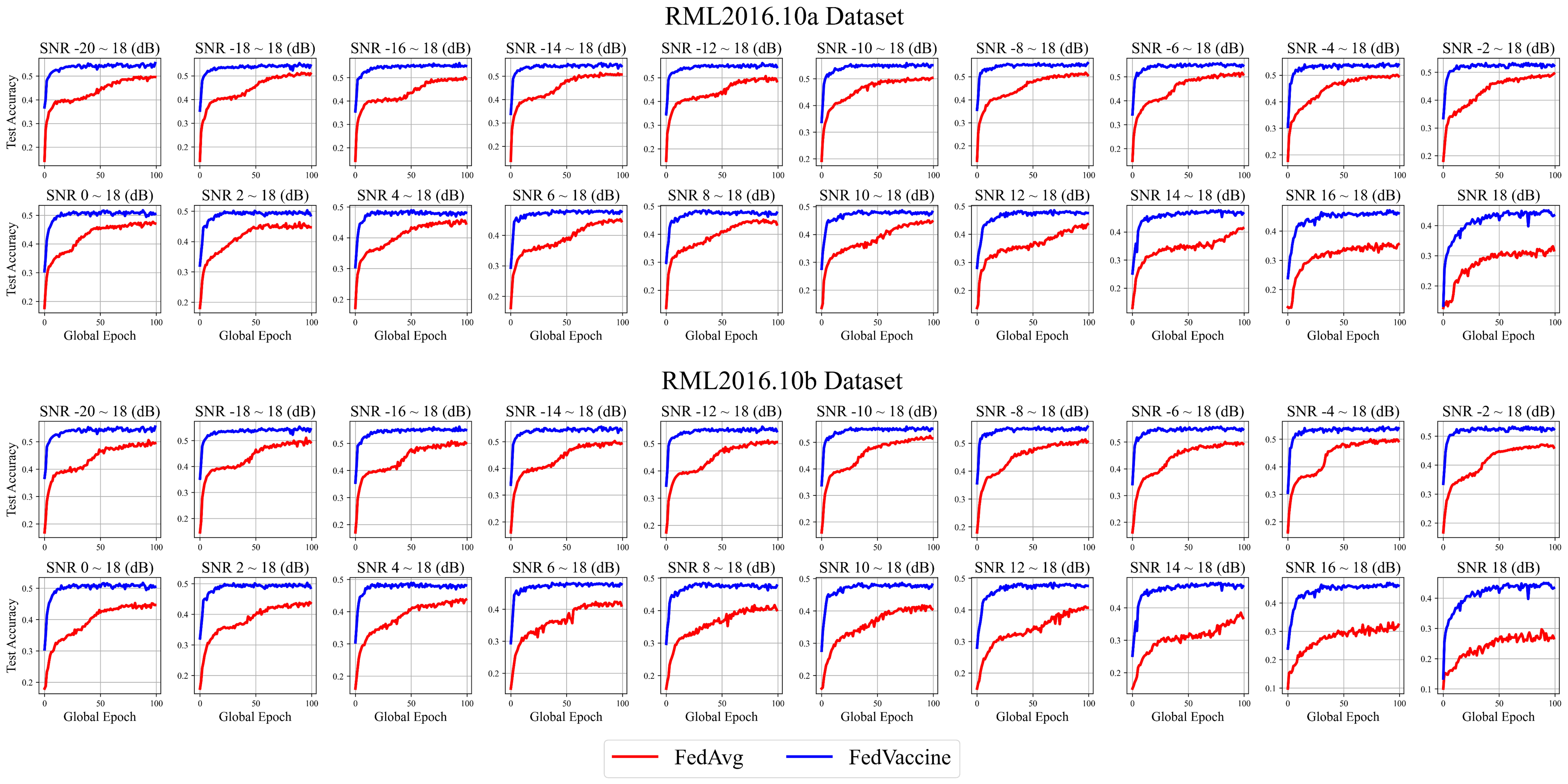}
    \caption{The test performance outcomes under IID conditions are compared between FedAvg and FedVaccine in each SNR range. It is noteworthy that the discernible performance gap widens as the training datasets encompass higher SNR values, culminating in disparities of approximately 12\% and 17\% for datasets RML2016.10a and RML2016.10b, respectively.}
    \label{fig:result_iid}
\end{figure*}

In the initial phase of our experiment, we undertake a comparative analysis of the fundamental performance between FedAvg, a baseline FL model, and the proposed FedVaccine within an IID environment across various SNR intervals. Fig.~\ref{fig:result_iid} visually represents the performance contrast within each SNR range of the training datasets, juxtaposing FedAvg and FedVaccine across both datasets.
Remarkably, FedVaccine demonstrates superior performance relative to FedAvg, manifesting accelerated convergence and exhibiting an outcome of achieving higher accuracy. To elucidate the quantitative disparities, Table~\ref{tab:iid_result} presents a comparative analysis of the maximal performance attained by the global models over 100 epochs. The FedVaccine performance was indicated to be highest in the SNR range between -12 to 18, whereas the FedAvg was -8 $\sim$ -10 to 18, with a slight difference between the peaks. Satisfying the equation~\eqref{eq:obj_harmonic_noise}, we set the $\theta$ to -12 dB in the non-IID experiments in FedVaccine.
Evidently, a discrepancy of 5 to 6\% in average performance is discernible between the two models, with disparities of 12\% and 17\% observed at their respective performance peaks for RML2016.10a and 10b datasets. Furthermore, the standard deviation associated with average performance underscores FedVaccine's propensity for stabilized convergence performance in contrast to FedAvg.

Additionally, this experiment substantiates the concept of harmonic noise resilience within the context of distributed learning environments, wherein the training datasets need not necessarily consist entirely of signals with high SNR. Instead, introducing a controlled degree of perturbation is shown to be advantageous for fostering robustness in the learning process.

\begin{table}[t]
\centering
\caption{The maximum test accuracy (with corresponding standard deviation) attained by FedAvg and FedVaccine under IID conditions is reported. The highest performances are highlighted in bold for clarity.}
\begin{tabular}{c|cccc}\hline
    & \multicolumn{2}{c}{RML2016.10a} & \multicolumn{2}{c}{RML2016.10b}  \\ \hline
    \toprule
    \textbf{SNR (dB)} & \textbf{FedAvg} & \textbf{FedVaccine} & \textbf{FedAvg} & \textbf{FedVaccine} \\
    \midrule
    -20 $\sim$ 18  & 49.82 & 55.49 & 50.60 & 55.40 \\
    -18 $\sim$ 18  & 51.32 & 55.38 & 50.73 & 55.74 \\
    -16 $\sim$ 18  & 50.18 & 56.20 & 50.85 & 56.10 \\
    -14 $\sim$ 18  & 51.45 & 55.61 & 50.26 & 56.16 \\
    -12 $\sim$ 18  & 50.71 & \textbf{56.21} & 51.20 & \textbf{56.48} \\
    -10 $\sim$ 18  & 50.35 & 55.90 & \textbf{52.42} & 56.33 \\
    -8 $\sim$ 18   & \textbf{51.80} & 55.96 & 51.37 & 56.14 \\
    -6 $\sim$ 18   & 51.74 & 55.76 & 50.06 & 55.76 \\
    -4 $\sim$ 18   & 50.12 & 54.57 & 49.98 & 54.41 \\
    -2 $\sim$ 18   & 49.56 & 53.30 & 47.13 & 53.28 \\
    0 $\sim$ 18 & 47.74 & 51.74 & 45.29 & 51.74 \\
    2 $\sim$ 18 & 46.38 & 50.30 & 44.00 & 50.20 \\ 
    4 $\sim$ 18 & 45.88 & 48.93 & 43.87 & 48.39 \\ 
    6 $\sim$ 18 & 45.30 & 48.23 & 42.34 & 47.72 \\ 
    8 $\sim$ 18 & 45.23 & 48.54 & 41.69 & 48.02 \\ 
    10 $\sim$ 18 & 45.05 & 48.70 & 41.75 & 47.56 \\ 
    12 $\sim$ 18 & 43.60 & 48.55 & 41.05 & 48.67 \\ 
    14 $\sim$ 18 & 41.37 & 47.70 & 38.51 & 47.42 \\
    16 $\sim$ 18 & 35.87 & 47.13 & 33.16 & 47.38 \\ 
    18 & 33.15 & 45.07 & 29.72 & 46.33 \\
    Average & 46.83(5.07) & 51.97(3.72) & 45.32(6.25) & 51.96(3.83) \\
    \bottomrule
\end{tabular}
\label{tab:iid_result}
\end{table}

\begin{figure*}
    \centering
    \includegraphics[width=\linewidth]{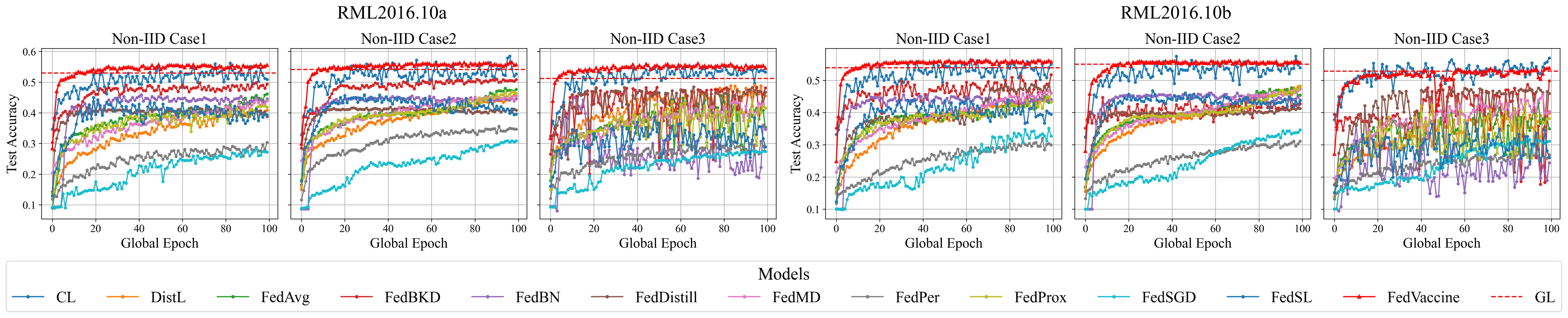}
    \caption{A comparative analysis of performance involving three distinct learning paradigms and a subset of FL models within three non-IID scenarios across two public datasets.}
    \label{fig:non-iid-result}
\end{figure*}

\subsection{Non-IID Results}

In this section, we investigate the effectiveness of FedVaccine across three prominent non-IID scenarios within the modulation classification task. The classification performance trajectories of various learning paradigms and FL models are depicted in Fig.~\ref{fig:non-iid-result}. Notably, the FedVaccine demonstrates higher performance compared to existing FL models in non-IID scenarios, achieving convergence at an accelerated pace. This outcome demonstrates the efficacy of serial learning in a non-IID environment, wherein FedVaccine successfully mitigates information loss during iterative aggregation stages as opposed to conventional aggregation processes. Furthermore, the discernment of an optimal SNR threshold contributes to robust performance, facilitating the vaccination effect. Additionally, the incorporation of a circulating dataset within the local queue enhances learning performance, further endorsing the efficacy of FedVaccine.

However, despite FedVaccine's expedited convergence, CL occasionally attains higher performance levels. This divergence can be attributed primarily to the larger dataset volumes fed into the CL model, which undergoes training with a much larger dataset volume within each global iteration. Notably, the conventional GL approach yields comparable performance, underscoring the learning efficacy of the traditional centralized learning approach. Remarkably, FedVaccine outperforms GL, thereby highlighting the effectiveness of our approach within non-IID contexts.

\subsection{Ablation Study} \label{ablation_study}


In this subsection, we conduct an ablation study to systematically analyze the contribution of individual components within the FedVaccine model by selectively modifying three parts: Cluster size, Queue size, and the SNR threshold $\theta$ range to discern their respective impacts on overall performance. The experimental protocol remained identical to the IID settings, where the performance variation was measured within the stabilized IID environment.

\subsubsection{Cluster Size}

In the iterative refinement of global models, determining the optimal cluster size represents a crucial hyperparameter in FedVaccine. In order to evaluate the efficacy of training across various cluster sizes, we categorize the cluster sizes into seven distinct configurations:

\begin{itemize}
    \item Cluster size 1, where the global model is updated using all local clients.
    
    \item Cluster size 2, wherein 50\% of the local clients are utilized per cluster, and the global model is updated twice within a global epoch.
    
    \item Cluster size 3, involving the utilization of approximately 33\% of the clients per cluster, with the global model being updated three times per global iteration. 
    
    \item Cluster size 4, using 25\% of clients per cluster, updating global model four times. 
    
    \item Cluster size 5, using 20\% of clients per cluster, updating global model five times.
    
    \item Cluster size 10, using 10\% of clients per cluster, updating global model ten times.
    
    \item A scenario without clustering, whereby local parameters are sequentially transmitted to the next local until all participant locals have undergone knowledge transfer within a single global iteration.
\end{itemize}

In accordance with the defined cluster sizes, we proceed to implement the FedVaccine algorithm and evaluate its performance, the results of which are presented in Table~\ref{tab:cluster_result}. Analysis of these results reveals that the different cluster size exerts influence on the learning outcomes within the IID scenario, highlighting the optimal cluster size is three, with a slight performance increase of 1 to 2\%. 



\begin{table}[t]
\centering
\caption{The examination of FedVaccine using different cluster sizes. The performance was measured using a test dataset, selecting the maximum accuracy (\%).}
\begin{tabular}{c|cc|cc}\hline
    & \multicolumn{2}{c|}{\textbf{RML2016.10a}} & \multicolumn{2}{c}{\textbf{RML2016.10b}}  \\ \hline
    \textbf{Cluster size} & \textbf{Accuracy} & \textbf{Loss} & \textbf{Accuracy} & \textbf{Loss} \\
    \midrule
    1  & 56.28 & 1.52 & 56.90 & 1.24 \\
    2  & 56.34 & 1.49 & 57.08 & 1.13 \\
    3  & \textbf{56.51} & \textbf{1.46} & \textbf{57.94} & \textbf{1.06} \\
    4  & 56.07 & 1.49 & 56.93 & 1.08 \\
    5  & 55.61 & 1.51 & 56.29 & 1.14 \\
    10 & 55.50 & 1.57 & 56.30 & 1.11 \\
    None & 55.96 & 1.66 & 56.12 & 1.12 \\
    \bottomrule
\end{tabular}
\label{tab:cluster_result}
\end{table}

\subsubsection{Queue Size}

The incorporation of a queue within our FedVaccine model serves to facilitate convergence during training, particularly in scenarios characterized by non-IID datasets, where biases may significantly impact training dynamics. This queue mechanism supplements the inherent degree of IID within the training dataset, assuming a memory capacity denoted by \(\vartheta\). In our experimental setup, based on the assumption of floating-point numbers represented with 4 bytes and sample shapes of (2, 128), we estimate that storing 1000 samples requires \(1024 \times 1000\) bytes, equivalent to 1000 KB. With a queue size of 1, representing the size of 1000 samples, \(\vartheta\) is set to 1000 KB. Table~\ref{tab:queue_size} presents the test accuracy and loss performance alongside the memory size of the queue across two modulation datasets. Although the performance appears unaffected by the queue size in these instances where the training datasets are IID, its indispensability becomes evident in unpredictable non-IID scenarios, emphasizing its role in ensuring robustness during training.


\begin{table}
\centering
\caption{The evaluation of FedVaccine's test performance across varying queue sizes, along with their respective memory requirements. Note that `Acc' in the table refers to accuracy (\%).}
\begin{tabular}{c|ccc|ccc}
\hline
\multicolumn{1}{l}{} & \multicolumn{3}{c|}{\textbf{RML2016.10a}} & \multicolumn{3}{c}{\textbf{RML2016.10b}} \\ \hline
\textbf{Queue} & \textbf{Acc} & \textbf{Loss} & \textbf{Memory} & \textbf{Acc} & \textbf{Loss} & \textbf{Memory} \\ \hline
None & 54.68 & 1.54 & Default (d) & 56.57 & 1.18 & Default (d) \\ 
1    & 54.73 & 1.46 & d+1000KB & 55.83 & 1.19 & d+1000KB \\
2    & 54.98 & 1.46 & d+2000KB & 55.93 & 1.19 & d+2000KB \\
3    & 55.43 & 1.47 & d+3000KB & 56.02 & 1.18 & d+3000KB \\
4    & 55.38 & 1.42 & d+4000KB & 56.18 & 1.18 & d+4000KB \\
5    & 55.57 & 1.40 & d+5000KB & 56.25 & 1.18 & d+5000KB \\
10   & 55.22 & 1.51 & d+10000KB & 55.90 & 1.17 & d+10000KB \\ \hline
\end{tabular}
\label{tab:queue_size}
\end{table}

\subsubsection{SNR Range}

As indicated in the previous section~\ref{sec:6-3}, the optimal SNR range (threshold $\theta$) was identified as -12 to 18 dB for FedVaccine. 
In the SNR ablation study, we partition the SNR range of the training datasets into four subsets: -20 to -10, -10 to 0, 0 to 10, and 10 to 18, without incorporating the highest SNR value, but dividing the range into four SNR levels. Utilizing this training set, we evaluate test performance across the entire SNR spectrum.

As shown in Table~\ref{tab:SNR_range_result}, performance within the -20 to -10 SNR range suggests poor trainability, with accuracy levels approximating random probability. Notably, while accuracy performance peaks within the SNR range of 0 to 9, corresponding loss values begin to diverge. Conversely, the SNR range of -10 to -1 yields the lowest loss scores, accompanied by similar accuracy levels observed within the 0 to 9 SNR range. These findings align with the results in section~\ref{section 5} in that certain noise levels propel the trainability. It demonstrates the SNR range of -10 to -1 as the optimal range for training the modulation classification model, where datasets exceeding SNR 0 demonstrate signs of overfitting, compromising generalizability.

\begin{table}[t]
\centering
\caption{The examination of FedVaccine's test performance across varying ranges of SNR. This ablation analysis provides insights into how the model performs across different SNR ranges, informing its robustness of the SNR range of -10 to -1 dB. Note that the unit of accuracy is \%.}
\begin{tabular}{c|cc|cc}\hline
    & \multicolumn{2}{c|}{\textbf{RML2016.10a}} & \multicolumn{2}{c}{\textbf{RML2016.10b}}  \\ \hline
    \textbf{SNR (dB)} & \textbf{Accuracy} & \textbf{Loss} & \textbf{Accuracy} & \textbf{Loss} \\
    \hline
    -20 $\sim$ -11 & 10.02 & 2.30 & 10.07 & 2.30 \\
    -10 $\sim$ -1  & 47.86 & \textbf{1.64} & 49.76 & \textbf{1.49} \\
    0 $\sim$ 9  & \textbf{50.01} & 4.76 & \textbf{50.23} & 2.36 \\
    10 $\sim$ 18 & 47.54 & 6.58 & 47.76 & 5.12 \\
    \bottomrule
\end{tabular}
\label{tab:SNR_range_result}
\end{table}


\section{Discussion} \label{sec7}

The following section elucidates the primary findings derived from our research on harmonic noise resilience and FedVaccine methodology for modulation classification. Emphasizing aspects of generalizability and practicality, we discuss the technical innovation and benefits intrinsic to our approach. Subsequently, we scrutinize the practical ramifications of these advancements and discuss the limitations of our study, along with prospective avenues for augmenting AMC within the domain of wireless communication applications.

\subsection{Harmonic Noise Resilience and Real-World Significance}

The findings from our harmonic noise resilience methodology in Sections \ref{section 5} and \ref{section 6} reveal noteworthy insights into the nuanced relationship between signal quality and noise levels in real-world applications. 
Notably, we demonstrated that optimal recognition performance does not consistently originate from low-noise signals; \textit{rather}, a delicate balance between signal fidelity and noise tolerance emerges as the key determinant of performance efficacy, as presented in Table~\ref{Tab:one_model_snr_acc} and Table~\ref{tab:iid_result}. 
Our harmonic noise resilience approach showed a new aspect of exploring the equilibrium between the original signal and inevitable noise sources, achieving the best modulation classification performance by learning regularized and balanced features across signals imbued with noise.
Our exhaustive experimentation underscores the efficacy of a novel approach to harmonic noise resilience, wherein an equilibrium is strategically forged between the intrinsic signal and the pervasive noise sources.
Our approach achieved the best modulation performance by learning regularized and balanced features across signals imbued with noise.

Beyond its implications for modulation classification, the concept of harmonic noise resilience holds promising implications for a plethora of machine learning-based recognition fields within wireless communication. These include channel estimation~\cite{soltani2019deep}, spectrum sensing~\cite{gao2019deep}, wireless security~\cite{sagduyu2019iot}, as well as location estimation and handover predictions~\cite{lee2020prediction}. The delineation of the intricate boundaries controlling the SNR heralds a paradigm shift in the conceptualization and deployment of wireless communication systems, thereby unlocking various untapped potentials.



\subsection{Technical Novelty and Advantages}


Existing FL-based AMC models have primarily targeted specific non-IID challenges, notably class imbalance, without possessing the requisite generalizability to address a spectrum of heterogeneous non-IID issues. Furthermore, the prevalent linear integration methodologies often entail information loss, thereby presenting formidable obstacles in constructing a truly effective global model.

In response to these challenges, our study introduces the FedVaccine model, tailored to confront the diverse non-IID challenges inherent in distributed signal environments, synergistically amalgamated with the harmonic noise resilience method.
The FedVaccine framework showcases resilience in handling signals plagued by intrinsic noise distortions, adeptly discerning robust features to augment the model's generalizability in real-world scenarios.
Additionally, our sequential model updates via segmenting the holistic parallel learning process into intra-cluster parallelism and inter-cluster serial learning, we mitigate information loss while amalgamating heterogeneous models.
Moreover, the adaptive queue storage propels the efficiency of fine-tuning the global model. 
Overall, our comprehensive experimentation corroborates the superior efficacy of the FedVaccine framework, affirming its proficiency in addressing the intricacies of distributed learning for modulation classification.

These notable advantages of \textit{robustness against noise} and \textit{enhanced generalizability in practical scenarios} position our FedVaccine model as a seminal advancement that bolsters its applicability within the domain of wireless communication.



\subsection{Limitations and Future Directions}

While our approach outperforms existing FL-based AMC methods in the non-IID domain, it grapples with a fundamental limitation of achieving significant performance across signals with a wide range of noises. 
Specifically, it still struggles to discern modulation signals amidst significantly high levels of noise. 
This challenge arises from the model's inability to differentiate between noise and the core essence of the signal, where we fundamentally leveraged features extracted within the signal merged with noise spectrums. 

To effectively train the AMC model to react within the variability of the noise signal, our interest lies in exploring a prototype learning approach that can make accurate predictions when encountered with unknown features.
By leveraging the representative features of modulation signals, prototype learning captures the essential nature of the modulation target. It comparatively measures the similarity between the representative feature and the sample-wise features within the feature space, where we believe it will enable an effective strategy to discern noise and modulation signals.

\section{Conclusion} \label{sec8}

In this study, we introduce FedVaccine, a novel Federated Learning framework tailored for modulation classification in wireless communication systems. The pervasive noise inherent in modulation signals poses a notable challenge to AI-driven distributed learning systems, hindering the optimization and practicality of classification models. Compounding this challenge are the dynamic non-IID attributes present across distributed datasets and temporal axes, impeding the conventional linear aggregation optimization process employed by FL methodologies and leading to information loss.

Our FedVaccine addresses these challenges through two main strategies. Firstly, we foster model robustness by intentionally exposing it to a balanced level of noise, which regularizes the training effect that mitigates overfitting. This optimal noise level is determined through our harmonic noise resilience approach and rigorously validated through extensive experimentation, demonstrating an enhanced level of generalizability across a diverse spectrum of SNRs.
Secondly, our framework significantly addresses the issue of non-IID attributes by partitioning the update process into distinct cluster sets, enabling multiple refinement of the global model through intra-cluster parameter aggregation and subsequent global model updates across inter-cluster iterations. Additionally, the incorporation of a dynamic queue structure within local devices facilitates adaptive dataset refreshing, thereby reducing bias and enhancing overall performance.

Our comprehensive experimental evaluations demonstrate that the FedVaccine outperforms existing FL models and several traditional learning paradigms in non-IID scenarios pertaining to modulation classification. These findings underscore the efficacy of FedVaccine in practical modulation classification systems within wireless networks. By offering a robust strategy to mitigate noise and address non-IID attributes, FedVaccine significantly advances the development of modulation classification systems, paving the way for more effective and reliable communication systems in practical deployment scenarios.

\section{Acknowledgement}


This research was supported by the MSIT (Ministry of Science and ICT), Korea, under the ITRC(Information Technology Research Center) support program (IITP-2024-RS-2024-00438056, 50\%) supervised by the IITP (Institute for Information \& Communications Technology Planning \& Evaluation).
This work was also supported by the Institute of Information \& communications Technology Planning \& Evaluation (IITP) grant funded by the Korean government (MSIT) (No. RS-2024-00396797, 50\%, Development of core technology for intelligent O-RAN security platform).

\bibliography{bib_file}
\bibliographystyle{IEEEtran}

\newpage




\end{document}